\documentclass[italian,english]{article}
\usepackage[T1]{fontenc}
\usepackage[latin9]{inputenc}
\usepackage{color}
\usepackage{graphicx}
\usepackage{esint}
\usepackage{babel}
\begin{document}

\title{\textbf{Interpretation of Mössbauer experiment in a rotating system:
a new proof for general relativity}}

\author{\textbf{Christian Corda}}

\maketitle
\begin{center}
Dipartimento di Scienze, Sezione di Fisica, Scuola Superiore di Studi
Universitari e Ricerca \textquotedbl{}Santa Rita\textquotedbl{}, via
San Nicola snc, 81049, San Pietro Infine (CE) Italy
\par\end{center}

\begin{center}
Austro-Ukrainian Institute for Science and Technology, Institut for
Theoretish Wiedner Hauptstrasse 8-10/136, A-1040, Wien, Austria
\par\end{center}

\begin{center}
International Institute for Applicable Mathematics \& Information
Sciences (IIAMIS),  B.M. Birla Science Centre, Adarsh Nagar, Hyderabad
- 500 463, India 
\par\end{center}

\begin{center}
\textit{E-mail address:} \textcolor{blue}{cordac.galilei@gmail.com} 
\par\end{center}
\begin{abstract}
A historical experiment by Kündig on the transverse Doppler shift
in a rotating system measured with the Mössbauer effect (Mössbauer
rotor experiment) has been recently first re-analyzed and then replied
by an experimental research group. The results of re-analyzing the
experiment have shown that a correct re-processing of Kündig's experimental
data gives an interesting deviation of a relative redshift between
emission and absorption resonant lines from the standard prediction
based on the relativistic dilatation of time. That prediction gives
a redshift $\frac{\nabla E}{E}\simeq-\frac{1}{2}\frac{v^{2}}{c^{2}}$
where $v$ is the tangential velocity of the absorber of resonant
radiation, $c$ is the velocity of light in vacuum and the result
is given to the accuracy of first-order in $\frac{v^{2}}{c^{2}}$.
Data re-processing gave $\frac{\nabla E}{E}\simeq-k\frac{v^{2}}{c^{2}}$
with $k=0.596\pm0.006$. Subsequent new experimental results by the
reply of Kündig experiment have shown a redshift with $k=0.68\pm0.03$
instead. 

By using Einstein Equivalence Principle, which states the equivalence
between the gravitational \textquotedbl{}force\textquotedbl{} and
the \emph{pseudo-force} experienced by an observer in a non-inertial
frame of reference (included a rotating frame of reference) here we
re-analyze the theoretical framework of Mössbauer rotor experiments
directly in the rotating frame of reference by using a general relativistic
treatment. It will be shown that previous analyses missed an important
effect of clock synchronization and that the correct general relativistic
prevision in the rotating frame gives $k\simeq\frac{2}{3}$ in perfect
agreement with the new experimental results. Such an effect of clock
synchronization has been missed in various papers in the literature
with some subsequent claim of invalidity of relativity theory and/or
some attempts to explain the experimental results through ``exotic''
effects. Our general relativistic interpretation shows, instead, that
the new experimental results of the Mössbauer rotor experiment are
a new, strong and independent, proof of Einstein general relativity.

In the final Section of the paper we discuss an analogy with the use
of General Relativity in Global Positioning Systems.
\end{abstract}

\section{Introduction}

The Mössbauer effect (discovered by R. Mössbauer in 1958 \cite{key-14})
consists in resonant and recoil-free emission and absorption of gamma
rays, without loss of energy, by atomic nuclei bound in a solid. It
resulted and currently results very important for basic research in
physics and chemistry. In this work we will focus on the so called
Mössbauer rotor experiment. In this particular experiment, the Mössbauer
effect works through an absorber orbited around a source of resonant
radiation (or vice versa). The aim is to verify the relativistic time
dilation time for a moving resonant absorber (the source) inducing
a relative energy shift between emission and absorption lines. 

In a couple of recent papers \cite{key-1,key-2}, the authors first
re-anayzed in \cite{key-1} the data of a known experiment of Kündig
on the transverse Doppler shift in a rotating system measured with
the Mössbauer effect \cite{key-3}, and second, they carried out their
own experiment on the time dilation effect in a rotating system \cite{key-2}.
In \cite{key-1} they found that the original experiment by Kündig
\cite{key-3} contained errors in the data processing. A puzzling
fact is that, after correction of the errors of Kündig, the experimental
data gave the value \cite{key-1}

\begin{equation}
\frac{\nabla E}{E}\simeq-k\frac{v^{2}}{c^{2}},\label{eq: k}
\end{equation}
where $k=0.596\pm0.006$, instead of the standard relativistic prediction
$k=0.5$ due to time dilatation. The authors of \cite{key-1} stressed
that the deviation of the coefficient $k$ in equation (\ref{eq: k})
from $0.5$ exceeds by almost 20 times the measuring error and that
the revealed deviation cannot be attributed to the influence of rotor
vibrations and other disturbing factors. All these potential disturbing
factors have been indeed excluded by a perfect methodological trick
applied by Kündig \cite{key-3}, i.e. a first-order Doppler modulation
of the energy of $\gamma-$quanta on a rotor at each fixed rotation
frequency. In that way, Kündig's experiment can be considered as the
most precise among other experiments of the same kind {[}4\textendash{}8{]},
where the experimenters measured only the count rate of detected $\gamma-$quanta
as a function of rotation frequency. The authors of \cite{key-1}
have also shown that the experiment {[}8{]}, which contains much more
data than the ones in {[}4\textendash{}7{]}, also confirms the supposition
$k>0.5.$ Motived by their results in \cite{key-1} the authors carried
out their own experiment \cite{key-2}. They decided to repeat neither
the scheme of the Kündig experiment \cite{key-3} nor the schemes
of other known experiments on the subject previously mentioned above
{[}4\textendash{}8{]}. In that way, they got independent information
on the value of $k$ in equation (\ref{eq: k}). In particular, they
refrained from the first-order Doppler modulation of the energy of
$\gamma-$quanta, in order to exclude the uncertainties in the realization
of this method \cite{key-2}. They followed the standard scheme {[}4\textendash{}8{]},
where the count rate of detected $\gamma-$quanta $N$ as a function
of the rotation frequency $\nu$ is measured. On the other hand, differently
from the experiments {[}4\textendash{}8{]}, they evaluated the influence
of chaotic vibrations on the measured value of $k$ \cite{key-2}.
Their developed method involved a joint processing of the data collected
for two selected resonant absorbers with the specified difference
of resonant line positions in the Mössbauer spectra \cite{key-2}.
The result obtained in \cite{key-2} is $k=0.68\pm0.03$, confirming
that the coefficient $k$ in equation (\ref{eq: k}) substantially
exceeds 0.5. The scheme of the new Mössbauer rotor experiment is in
Figure 1, while technical details on it can be found in \cite{key-2}. 

\begin{figure}
\includegraphics[scale=0.75]{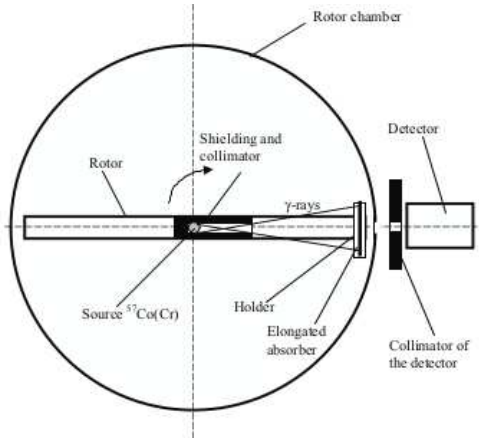}

\caption{Scheme of the new Mössbauer rotor experiment, adapted from ref. \cite{key-2}}
\end{figure}

In this paper, Einstein Equivalence Principle, which states the equivalence
between the gravitational \textquotedbl{}force\textquotedbl{} and
the \emph{pseudo-force} experienced by an observer in a non-inertial
frame of reference (included a rotating frame of reference) will be
used to re-analyze the theoretical framework of Mössbauer rotor experiments
directly in the rotating frame of reference by using a general relativistic
treatment. The results will show that previous analyses missed an
important effect of clock synchronization and that the correct general
relativistic prevision gives $k\simeq\frac{2}{3}$ in perfect agreement
with the new experimental results of \cite{key-2}. In that way, the
general relativistic interpretation of this paper shows that the new
experimental results of the Mössbauer rotor experiment are a new,
strong and independent proof of Einstein general relativity. We stress
that various papers in the literature missed the effect of clock synchronization
\cite{key-1}-\cite{key-8}, \cite{key-11}-\cite{key-13} with some
subsequent claim of invalidity of relativity theory and/or some attempts
to explain the experimental results through ``exotic'' effects \cite{key-1,key-2,key-11,key-12,key-13}.

\section{General relativistic interpretation of time dilatation}

Following \cite{key-9} let us consider a transformation from an inertial
frame, in which the space-time is Minkowskian, to a rotating frame
of reference. Using cylindrical coordinates, the line element in the
starting inertial frame is \cite{key-9} 

\begin{equation}
ds^{2}=c^{2}dt^{2}-dr^{2}-r^{2}d\phi^{2}-dz^{2}.\label{eq: Minkowskian}
\end{equation}
The transformation to a frame of reference $\left\{ t',r',\phi'z'\right\} $
rotating at the uniform angular rate $\omega$ with respect to the
starting inertial frame is given by \cite{key-9} 

\begin{equation}
\begin{array}{cccc}
t=t'\; & r=r' & \;\phi=\phi'+\omega t'\quad & z=z'\end{array}.\label{eq: trasformazione Langevin}
\end{equation}
Thus, eq. (\ref{eq: Minkowskian}) becomes the following well-known
line element (Langevin metric) in the rotating frame \cite{key-9}
\begin{equation}
ds^{2}=\left(1-\frac{r'^{2}\omega^{2}}{c^{2}}\right)c^{2}dt'^{2}-2\omega r'^{2}d\phi'dt'-dr'^{2}-r'^{2}d\phi'^{2}-dz'^{2}.\label{eq: Langevin metric}
\end{equation}
The transformation (\ref{eq: trasformazione Langevin}) is both simple
to grasp and highly illustrative of the general covariance of GR as
it shows that one can work first in a \textquotedbl{}simpler\textquotedbl{}
frame and then transforming to a more \textquotedbl{}complex\textquotedbl{}
one \cite{key-15}. As we consider light propagating in the radial
direction ($d\phi'=dz'=0$), the line element (\ref{eq: Langevin metric})
reduces to 

\begin{equation}
ds^{2}=\left(1-\frac{r'^{2}\omega^{2}}{c^{2}}\right)c^{2}dt'^{2}-dr'^{2}.\label{eq: metrica rotante}
\end{equation}
Einstein Equivalence Principle permits to interpret the line element
(\ref{eq: metrica rotante}) in terms of a curved spacetime in presence
of a static gravitational field. Setting the origin of the rotating
frame in the source of the emitting radiation, we have a first contribution
which arises from the ``gravitational redshift'' that can be directly
computed using eq. (25.26) in \cite{key-10} , which in the twentieth
printing 1997 of \cite{key-10} is written as 
\begin{equation}
z\equiv\frac{\Delta\lambda}{\lambda}=\frac{\lambda_{received}-\lambda_{emitted}}{\lambda_{emitted}}=|g_{00}(r'_{1})|^{-\frac{1}{2}}-1\label{eq: z  MTW}
\end{equation}
and represents the redshift of a photon emitted by an atom a rest
an a gravitational field and received by an observer at rest at infinity.
Here we use a slightly different equation with respect to eq. (25.26)
in \cite{key-10} because here we are considering a gravitational
field which increases with increasing radial coordinate $r'$ while
eq. (25.26) in \cite{key-10} concerns a gravitational field which
decreases with increasing radial coordinate. Also, we set the zero
potential in $r'=0$ instead of at infinity and we use the proper
time instead of the wavelength $\lambda$. Thus, combining eq. (\ref{eq: metrica rotante}),
we get 
\begin{equation}
\begin{array}{c}
z_{1}\equiv\frac{\nabla\tau_{10}-\nabla\tau_{11}}{\tau}=1-|g_{00}(r'_{1})|{}^{-\frac{1}{2}}=1-\frac{1}{\sqrt{1-\frac{\left(r'_{1}\right)^{2}\omega^{2}}{c^{2}}}}\\
\\
=1-\frac{1}{\sqrt{1-\frac{v^{2}}{c^{2}}}}\simeq-\frac{1}{2}\frac{v^{2}}{c^{2}},
\end{array}\label{eq: gravitational redshift}
\end{equation}
where $\nabla\tau_{10}$ is the delay of the emitted radiation, $\nabla\tau_{11}$
is the delay of the received radiation, $r'_{1}\simeq c\tau$ is the
radial distance between the source and the detector and $v=r'_{1}\omega$
is the tangential velocity of the detector. Hence, we find a first
contribution, say $k_{1}=\frac{1}{2}$, to $k$.

\section{Clock synchronization}

We stress that we calculated the variations of proper time $\nabla\tau_{10}$
and $\nabla\tau_{11}$ in the origin of the rotating frame which is
located in the source of the radiation. But the detector is moving
with respect to the origin in the rotating frame. Thus, the clock
in the detector must be synchronized with the clock in the origin,
and this gives a second contribution to the redshift. To compute this
second contribution we use eq. (10) of \cite{key-9} which represents
the proper time increment $d\tau$ on the moving clock having radial
coordinate $r'$ for values $v\ll c$ 

\begin{equation}
d\tau=dt'\left(1-\frac{r'^{2}\omega^{2}}{c^{2}}\right).\label{eq:secondo contributo}
\end{equation}
Inserting the condition of null geodesics $ds=0$ in eq. (\ref{eq: metrica rotante})
one gets 
\begin{equation}
cdt'=\frac{dr'}{\sqrt{1-\frac{r'^{2}\omega^{2}}{c^{2}}}},\label{eq: tempo 2}
\end{equation}
where we take the positive sign in the square root because the radiation
is propagating in the positive $r$ direction. Combining eqs. (\ref{eq:secondo contributo})
and (\ref{eq: tempo 2}) one obtains

\begin{equation}
cd\tau=\sqrt{1-\frac{r'^{2}\omega^{2}}{c^{2}}}dr'.\label{eq: secondo contributo finale}
\end{equation}
Eq. (\ref{eq: secondo contributo finale}) is well approximated by
\begin{equation}
cd\tau\simeq\left(1-\frac{1}{2}\frac{r'^{2}\omega^{2}}{c^{2}}+....\right)dr',\label{eq: well approximated}
\end{equation}
which permits to find the second contribution of order $\frac{v^{2}}{c^{2}}$
to the variation of proper time as 
\begin{equation}
c\nabla\tau_{2}=\int_{0}^{r'_{1}}\left(1-\frac{1}{2}\frac{\left(r'_{1}\right)^{2}\omega^{2}}{c^{2}}\right)dr'-r'_{1}=-\frac{1}{6}\frac{\left(r'_{1}\right)^{3}\omega^{2}}{c^{2}}=-\frac{1}{6}r'_{1}\frac{v^{2}}{c^{2}}.\label{eq: delta tau 2}
\end{equation}

Thus, as $r'_{1}\simeq c\tau$ is the radial distance between the
source and the detector, we get the second contribution of order $\frac{v^{2}}{c^{2}}$
to the redshift as 
\begin{equation}
z_{2}\equiv\frac{\nabla\tau_{2}}{\tau}=-k_{2}\frac{v}{c^{2}}^{2}=-\frac{1}{6}\frac{v^{2}}{c^{2}}.\label{eq: z2}
\end{equation}
Then, we obtain $k_{2}=\frac{1}{6}$ and using eqs. (\ref{eq: gravitational redshift})
and (\ref{eq: z2}) the total redshift is 
\begin{equation}
\begin{array}{c}
z\equiv z_{1}+z_{2}=\frac{\nabla\tau_{10}-\nabla\tau_{11}+\nabla\tau_{2}}{\tau}=-\left(k_{1}+k_{2}\right)\frac{v^{2}}{c^{2}}\\
\\
=-\left(\frac{1}{2}+\frac{1}{6}\right)\frac{v^{2}}{c^{2}}=-k\frac{v^{2}}{c^{2}}=-\frac{2}{3}\frac{v^{2}}{c^{2}}=0.\bar{6}\frac{v^{2}}{c^{2}},
\end{array}\label{eq: z totale}
\end{equation}
which is completely consistent with the result $k=0.68\pm0.03$ in
\cite{key-2}.

\section{Discussion of the correction due to clock synchronization and analogy
with the use of General Relativity in Global Positioning Systems }

We stress that the additional factor $-\frac{1}{6}$ in eq. (\ref{eq: z2})
comes from clock synchronization \cite{key-15}. In other words, its
theoretical absence in the works \cite{key-1}-\cite{key-8}, \cite{key-11}-\cite{key-13}
reflected the incorrect comparison of clock rates between a clock
at the origin and one at the detector \cite{key-15}. This generated
wrong claims of invalidity of relativity theory and/or some attempts
to explain the experimental results through ``exotic'' effects \cite{key-1,key-2,key-11,key-12,key-13}
which, instead, must be rejected.

We evoked the appropriate reference \cite{key-9} for a discussion
of the Langevin metric. This is dedicated to the use of General Relativity
in Global Positioning Systems (GPS), which leads to the following
interesting realisation \cite{key-15}: the correction of $-\frac{1}{6}$
in eq. (\ref{eq: z2}) is analog to the correction that one must consider
in GPS when accounting for the difference between the time measured
in a frame co-rotating with the Earth geoid and the time measured
in a non-rotating (locally inertial) Earth centered frame (and also
the difference between the proper time of an observer at the surface
of the Earth and at infinity). Indeed, if one simply considers the
Schwarzschild gravitational redshift but neglects the effect of the
Earth's rotation, GPS would not work \cite{key-15}! In fact, following
Chapters 3 and 4 of \cite{key-9} in order to address the problem
of clock synchronization within the GPS one starts from an approximate
solution of Einstein\textquoteright{}s field equations in isotropic
coordinates in a locally inertial, non-rotating, freely falling coordinate
system with origin at the earth's center of mass, i.e. eq. (12) in
\cite{key-9} which is 
\begin{equation}
ds^{2}=\left(1+\frac{2V}{c^{2}}\right)\left(cdt\right)^{2}-\left(1-\frac{2V}{c^{2}}\right)\left(dr^{2}+r^{2}d\theta^{2}+r^{2}\sin^{2}\theta d\phi^{2}\right)\label{eq: approximate solution}
\end{equation}
where $V$ is the Newtonian gravitational potential of the earth and
${r,\theta,\phi}$ are spherical polar coordinates. $V$ is given
approximately by \cite{key-9}: 
\begin{equation}
V\simeq\frac{-GM_{E}}{r}\left[1-J_{2}\left(\frac{a_{1}}{r}\right)^{2}P_{2}\cos\theta\right],\label{eq: V}
\end{equation}
where $M_{E}$ is the earth's mass, $J_{2}$ the earth's quadrupole
moment coefficient, $a_{1}$ the earth's equatorial radius and $P_{2}$
the Legendre polynomial of degree $2$ \cite{key-9}. Usually, one
retains only terms of first order in the small quantity $\frac{V}{c^{2}}$
in eq. (\ref{eq: approximate solution}), while higher multipole moment
contributions to eq. (\ref{eq: V}) have negligible effect for relativity
in GPS \cite{key-9}. The equivalent transformations of eqs. (\ref{eq: trasformazione Langevin})
for spherical polar coordinates are \cite{key-9}: 
\begin{equation}
t=t'\;.r=r'\;\theta=\theta'\;\phi'+\omega_{E}t'\label{eq: Langevin sferiche}
\end{equation}
where $\omega_{E}$ is the earth's uniform angular rate. Applying
the transformations (\ref{eq: Langevin sferiche}) to the line element
(\ref{eq: approximate solution}) and retaining only terms of order
$1/c^{2}$ one gets the line element for the so called earth-centered,
earth-fixed, reference frame (ECEF frame) \cite{key-9}: 
\begin{equation}
\begin{array}{c}
ds^{2}=\left[1+\frac{2V}{c^{2}}-\left(\frac{\omega_{E}r'\sin\theta'}{c}\right)^{2}\right]\left(cdt'\right)^{2}+2\omega_{E}r'^{2}\sin^{2}\theta'd\phi'dt'\\
\\
-\left(1-\frac{2V}{c^{2}}\right)\left(dr'^{2}+r'^{2}d\theta'^{2}+r'^{2}\sin^{2}\theta'd\phi'^{2}\right).
\end{array}\label{eq: ECEF frame}
\end{equation}
 The rate of coordinate time in eq. (\ref{eq: approximate solution}
is defined by standard clocks at rest at infinity \cite{key-9}. We
prefer to consider the rate of coordinate time by standard clocks
at rest on the earth's surface \cite{key-9}. Then, a new coordinate
time $t''$ can be defined through a constant rate change \cite{key-9}:
\begin{equation}
t''=\left(1+\frac{\Phi_{0}}{c^{2}}\right)t'=\left(1+\frac{\Phi_{0}}{c^{2}}\right)t,\label{eq: constant rate change}
\end{equation}
where \cite{key-9} 
\begin{equation}
\Phi_{0}\equiv-\frac{GM_{E}}{a_{1}}-\frac{GM_{E}J_{2}}{2a_{1}}-\frac{1}{2}\left(\omega_{E}a_{1}\right)^{2},\label{eq: fi zero}
\end{equation}
and the correction (\ref{eq: constant rate change}) is order seven
parts in $10^{10}$ \cite{key-9}. Applying this time scale change
in the ECEF line element (\ref{eq: ECEF frame}) and reteining only
terms of order $\frac{1}{c^{2}}$ we obtain \cite{key-9} 
\begin{equation}
\begin{array}{c}
ds^{2}=\left[1+\frac{2\left(\Phi-\Phi_{0}\right)}{c^{2}}\right]\left(cdt''\right)^{2}+2\omega_{E}r'^{2}\sin^{2}\theta'd\phi'dt''\\
\\
-\left(1-\frac{2V}{c^{2}}\right)\left(dr'^{2}+r'^{2}d\theta'^{2}+r'^{2}\sin^{2}\theta'd\phi'^{2}\right),
\end{array}\label{eq: ECEF frame 2}
\end{equation}
where \cite{key-9} 
\begin{equation}
\Phi\equiv\frac{V}{c^{2}}-\frac{1}{2}\left(\frac{\omega_{E}r'\sin\theta'}{c}\right)^{2}.\label{eq: fi}
\end{equation}
is the effective gravitational potential in the rotating frame. Applying
the time scale change (\ref{eq: constant rate change}) in the non-rotating
line element (\ref{eq: approximate solution}) and dropping the primes
on $t''$ in order to just use the symbol $t$ one gets \cite{key-9}
\begin{equation}
ds^{2}=\left[1+\frac{2\left(V-\Phi_{0}\right)}{c^{2}}\right]\left(cdt\right)^{2}-\left(1-\frac{2V}{c^{2}}\right)\left(dr^{2}+r^{2}d\theta^{2}+r^{2}\sin^{2}\theta d\phi^{2}\right).\label{eq: approximate solution rescaled}
\end{equation}
The coordinate time $t$ in eq. (\ref{eq: approximate solution rescaled})
is valid in a very large coordinate patch which covers both the earth
and the GPS satellite constellation \cite{key-9}. Thus, such a coordinate
time is used as a basis for synchronization in the earth's neighborhood
\cite{key-9}. The difference $\left(V-\Phi_{0}\right)$ in the first
term of eq. (\ref{eq: approximate solution rescaled}) is due to the
issue that in the underlying earth-centered locally inertial frame
in which the line element (\ref{eq: approximate solution rescaled})
is expressed, the unit of time is determined by moving clocks in a
spatially-dependent gravitational field \cite{key-9}. We observe
that eq. (\ref{eq: approximate solution rescaled}) contains both
the effects of apparent slowing of moving clocks and frequency shifts
due to gravitation \cite{key-9}. This implies that the proper time
elapsing on the orbiting GPS clocks cannot be simply used to transfer
time from one transmission event to another because path-dependent
effects must be taken into due account, exactly like in the discussion
of clock synchronization in Section 3.

The discussion in this Section has been inserted in order to highlight
that the obtained correction $-\frac{1}{6}$ in eq. (\ref{eq: z2})
is not an obscure mathematical or physical detail, but a fundamental
ingredient that must be taken into due account \cite{key-15}.

\section{Conclusion remarks}

In this paper Einstein Equivalence Principle, stating the equivalence
between the gravitational \textquotedbl{}force\textquotedbl{} and
the \emph{pseudo-force} experienced by an observer in a non-inertial
frame of reference (included a rotating frame of reference) has been
used to re-analyze the theoretical framework of the new Mössbauer
rotor experiment in \cite{key-2} directly in the rotating frame of
reference by using a general relativistic treatment. The results have
shown that previous analyses missed an important effect of clock synchronization
and that the correct general relativistic prevision gives $k\simeq\frac{2}{3}$
in perfect agreement with the new experimental results in \cite{key-2}.
Thus, the general relativistic interpretation in this letter shows
that the new experimental results of the Mössbauer rotor experiment
are a new, strong and independent proof of Einstein general relativity.
The importance of our results is stressed by the issue that various
papers in the literature missed the effect of clock synchronization\cite{key-1}-\cite{key-8},
\cite{key-11}-\cite{key-13} with some subsequent claim of invalidity
of relativity theory and/or some attempts to explain the experimental
results through ``exotic'' effects \cite{key-1,key-2,key-11,key-12,key-13}.

An analogy with the use of General Relativity in Global Positioning
Systems has been highlight in the final Section of the paper.

\subsection*{Acknowledgements }

It is a pleasure to thank Alexander Kholmetskii for pointing out to
me problems in Mössbauer rotor experiment and Ugo Abundo for discussions
on the the arguments of this letter. The author thanks the anonimous
referees for the important points they raised which permitted to strongly
improve this paper.

\end{document}